\documentstyle[11pt]{article}
\newcommand{\be}{\begin{equation}}
\newcommand{\ee}{\end{equation}}
\newcommand{\ba}{\begin{eqnarray}}
\newcommand{\ea}{\end{eqnarray}}
\newcommand{\bann}{\begin{eqnarray*}}
\newcommand{\eann}{\end{eqnarray*}}

\begin{document}
\hbadness=10000

\title{{\bf \huge  Pre-Equilibrium Stage
and Phase Transition of Quark Matter Probed by Photon Interferometry
\footnote[0]{\Large Submitted to Physical Review Letters}
}}
\vspace{-3.0cm}
{\huge \bf }
\author{
U. Ornik$^1$\thanks{E.Mail: ORNIK@TPRI6O.GSI.DE},
M. Pl\"umer$^2$\thanks{E. Mail: PLUEMER@MAILER.UNI-MARBURG.DE},
A. Timmermann$^3$\thanks{E. Mail: TIMMERMANN@DKRZ.D400.DE}{ }
 and R.M.
Weiner$^2$\thanks{E. Mail: WEINER@MAILER.UNI-MARBURG.DE} }

\date{$^1$ GSI, Darmstadt, FRG\\
$^2$ Physics Department, Univ. of Marburg, Marburg, FRG\\
$^3$ Max-Planck Institute of Meteorology, Hamburg, FRG }

\thispagestyle{empty}

\maketitle

\vspace{0.5cm}

\begin{abstract}
We study single- and double-inclusive spectra of thermal photons,
produced in heavy-ion collisions at $\sqrt{s}=200$ AGeV within a
realistic
space-time framework which combines the Parton-Cascade-Model and
3-dimensional hydrodynamics (HYLANDER). This allows also for the
first time to take into account pre-equilibrium effects
for photon production. A rapid decrease in the width of the
correlation function as the photon transverse momentum drops below
$\sim 1.5$ GeV is a signature of the deconfinement phase transition.
\end{abstract}
\newpage

Under extreme conditions -- expected to
 be realized, e.g., in high energy heavy ion collision experiments
to be performed at the Relativistic Heavy-Ion Collider (RHIC) --
the theory of strong interaction (quantum chromodynamics) predicts
a phase transition from hadronic matter to a state of
 quasi-free quarks and gluons  - the quark-gluon plasma (QGP).
The detection of this transition is one of the most important and debated
problems of high energy nuclear physics.
The problem is the development of
a suitable strategy for distinguishing between a system  in which
a QGP was formed at some early or intermediate stage of its evolution
and a system which consisted of purely hadronic matter
at all times. Two types of probes have been proposed for this
strategy: hadronic probes and electromagnetic ones. Hadronic signals
have two serious disadvantages:\\
(a) They appear only in the last stage of evolution of the system
while the QGP, if it has been formed at all, exists only at the
beginning of its evolution.\\
(b) There exists no satisfactory method yet to treat strong
interactions in the soft (non-perturbative QCD)
regime.\\
Electromagnetic probes on the other hand,
such as direct photons and dileptons,
due to their large mean free paths and small cross sections
pass through the plasma with very little distortion from
strong and electromagnetic interactions. Thus, photons and dileptons
retain the
memory of the early stages of the  evolution \cite{shuryak10}.

Our aim is to study inclusive one- and two-photon spectra
at the RHIC energy, $\sqrt{s}=200$ AGeV,
in a realistic 3-dimensional framework.
Particularly Bose-Einstein correlations (BEC) for thermal photons
are very sensitive to the thermal history and the space-time
development of the source.
In a previous 1-dimensional calculation \cite{ich}, it was pointed out
that the
transverse momentum dependence of the
longitudinal correlation functions probes the
deconfinement phase transition.
A rapid change in the shape of the correlation function from
a two component to a single component structure was observed.
This change is directly related to the first order phase transition
from the QGP to a hadronic phase.
The sensitivity of BECs
to a phase transition is also considered in Ref.\cite{longevity}
where a small qualitative change of the transverse
correlation functions was observed.
However,  the observation of this effect would require
a very high
experimental sensitivity
of the correlation function of the order $10^{-3}-10^{-4}$
which is quite unrealistic at present.

 All calculations which have been performed so far
\cite{ich,longevity,history,fidelity} are lacking  a realistic
space-time description of the source, ignore the
pre-equilibrium stage and assume
chemical equilibrium even in the
early stages of the  evolution.  Moreover, previous studies
\cite{ich,longevity,history,fidelity}
of BEC's of thermal photons used expressions for the rates of photons
emitted from the hadronic system which did not take into account the
contribution \cite{xiong} from the $A_1$ meson
and thus considerably
underestimated the rates.

In order to improve this situation we include several new features
for the description of the photon production at RHIC.
For the space-time evolution we adopt
a hybrid model, which implements an exact and fully 3-dimensional
hydrodynamical simulation (HYLANDER \cite{jan}) and features of the
Parton-Cascade-Model (PCM) \cite{geiger,geiger1,geiger2,geigerchem}.
For the photon production rates in the quark-gluon phase, the mixed phase
and the hadronic phase the same expressions were used as in
Ref.\cite{ich1}.

 Full local thermaliztion is not reached
 instantaneously in ultrarelativistic heavy-ion collisions.
Non-equilibrium models
\cite{geiger,wang}
which take into account parton-parton interactions  in the
early stages of the collision
suggest that local thermal
equilibrium is reached after some time on the order
of $\sim 0.1-2$ fm/c.
For high energies, as will be reached, e.g., in heavy-ion collisions
at RHIC, this type of scenario is generally accepted as the most
realistic description presently available. Thus it becomes necessary to
consider
photon emission in the absence of chemical equilibrium.
This was done in
Ref.\cite{strickland} and we shall apply this procedure below.
 The fugacities
$\lambda_q, \lambda_{\overline{q}}, \lambda_g$
for quarks and gluons are introduced via the statistical distributions
of the particles:
\begin{equation}
f_{q,\overline{q}} = \frac{\lambda_{q,\overline{q}}}
{\lambda_{q,\overline{q}}+e^{p_0/T}}
\label{eq:qfugazi}
\end{equation}
and for gluons analogously, replacing the $+$ by $-$. Here $p_0$ denotes
the energy in the rest frame of the particle.
The expressions for the photon rates
of the Compton and annihilation process in a non-equilibrium plasma
are given as \cite{strickland}\\
{\it (Compton:)}
\begin{equation}
E_{\gamma} \frac{dN}{d^3kd^4x} =
\frac{5}{9} \frac{2 \alpha \alpha_s}{\pi^4} \lambda_q \lambda_g T^2
e^{-E/T} \sum_{n=0}^\infty \frac{(-\lambda_q)^n}{(n+1)^2}
\left[\ln \left(\frac{12E}{g^2 \kappa^2T(n+1)}\right)+\frac{1}{2}
-C \right]
\label{eq:noneqcom}
\end{equation}
and\\
{\it (Annihilation:)}
\begin{equation}
E_{\gamma} \frac{dN}{d^3kd^4x} =
\frac{5}{9} \frac{2 \alpha \alpha_s}{\pi^4} \lambda_q \lambda_{\overline{q}}
T^2
e^{-E/T} \sum_{n=0}^\infty \frac{{\lambda_g}^n}{(n+1)^2}
\left[\ln \left(\frac{12E}{g^2 \kappa^2T(n+1)}\right)-1
-C \right] ,
\label{eq:noneqanni}
\end{equation}
where $g=\sqrt{4\pi\alpha_s}$ and $C=0.57721...$ .
In Ref. \cite{strickland} $\kappa$ is considered to be a function of
$\lambda_g$ and $\lambda_q$.
We shall put $\kappa$ to 1, which corresponds
to a simplified
thermal quark mass of $m_q^2=\frac{1}{6}g^2 T^2$.\footnote{
Note, that the rates in (\ref{eq:noneqanni})
diverge, if $\lambda_g>1$, which
corresponds to a gluonic condensate.}

In our model
the space-time evolution of $Au+Au$ collisions at $\sqrt{s}=200$ AGeV
can be divided into two stages.

{\bf 1.} The first stage is associated with the pre-equilibrium dynamics,
which is governed by hard and semihard QCD processes.
The development of the initial nuclear parton distributions
is calculated within  the PCM
\cite{geiger,geiger1,geiger2,geigerchem}.
It is found \cite{geiger} that for $Au+Au$ collisions at RHIC kinetical
equilibrium is established for quarks and
for gluons at the latest after $2$ fm/c.
Gluons reach  both chemical and thermal (local) equilibrium.
Chemical equilibration of quarks, on the other hand,
 takes much longer
as the chemical equilibration time for light quarks is too  large
 to compete
with the dilution of the expanding system.
The fugacity of light quarks e.g. reaches only a value of $\sim 0.6$
instead of $1$ as would correspond to full equilibrium
\cite{geigerchem}.
Actually, the equilibrium QGP at the RHIC experiment is dominated by
gluons.

{\bf 2.}  As soon as the hot and dense system
has reached local thermal equilibrium it is more realistic to describe
the further evolution of the fireball
within a hydrodynamical model.

The equations of relativistic hydrodynamics
are solved exactly and fully 3-dimensionally \cite{udo,jan}
with the numerical code HYLANDER.
In our calculation we adopt an equation of state (EOS)
which was obtained
by lattice-gauge simulations \cite{redlich}.
The principal ingredients of hydrodynamics are the
initial conditions which are obtained directly from
the PCM.
At the ``initial time''
$t_i=2.4$ fm/c
we extract the velocity field, the baryon-density distributions and the
energy-density profile from the PCM and use
it in the hydrodynamical calculation.

In order to obtain the single-inclusive spectra
for thermal photons\footnote{
This includes also the pre-equilibrium photons,
calculated with Eq.(\ref{eq:noneqcom}),(\ref{eq:noneqanni}).}
it is necessary
to integrate the photon rates over the space-time region defined by
the space-time evolution and the
thermal and chemical history of the hot and dense matter.
The fugacities, the temperature field and the four velocity field
are obtained in the pre-equilibrium stage from the PCM
and afterwards from hydrodynamics.

Fig. 1  shows the separate contributions of (i) the pre-equilibrium stage,
(ii) the equilibrated QGP phase, (iii) the mixed phase and
(iv) the hadronic phase to the
single-inclusive spectra of thermal photons for
the reaction $Au+Au$ at $\sqrt{s}=$ 200 AGeV.
In the transverse momentum region
$k_\perp>1.5$ GeV the pre-equilibrium outshines all other phases.
This is due to the high
initial effective temperatures of $T_i\sim 950$ MeV in the PCM.
Only at $k_\perp\sim $1 GeV, the contribution from the purely hadronic
phase becomes comparable to that of the pre-equilibrium, while the
rate of photons emitted from the equilibrium QGP and from the mixed
cannot compete with the pre-equibrium yield in the transverse momentum
range considered here ($1$ GeV $\leq k_\perp \leq$ $3$ GeV).
Thus, only for photons within the transverse momentum
window $k_\perp \sim 1-2$ GeV there is a chance that the correlations
are sensitive to the presence of two separate phases: the pre-equilibrium
stage, and the purely hadronic phase.

In ultrarelativistic nuclear and particle collision experiments
BEC have been measured almost exclusively for pions and kaons so far.
These particles decouple from the hydrodynamical system
on the freeze-out hypersurface and thus carry only information on
the late stage of the evolution.
BEC for thermal photons on the other hand are sensitive to the entire
space-time evolution of the system.
They carry information on the pre-equilibrium QGP,
the fully equilibrated QGP, the phase transition
and the hadronic matter.

We consider the second order correlation function
$C_2(\vec{k}_1,\vec{k}_2)$ for identical particles:
\begin{equation}
C_2(\vec{k}_1,\vec{k}_2) = \frac{P_2(\vec{k}_1,\vec{k}_2)}
{P_1(\vec{k}_1)P_1(\vec{k}_2)},
\end{equation}
where $\vec{k}_i \ (i=1,2)$ are the three momenta of the particles
 and $P_1(\vec{k})$  and $P_2(\vec{k}_1,\vec{k}_2)$
are the one- and two-particle inclusive distributions.

For the general case of a Gaussian density matrix and a totally chaotic
source the thermal Wick theorem holds and
the two-particle inclusive distribution can be written in the form
\cite{ich}
\begin{equation}
P_2(\vec{k}_1,\vec{k}_2) = P_1(\vec{k}_1)P_1(\vec{k}_2) + \left|
\int d^4x \ w\left(x,\frac{k_1+k_2}{2}\right)
\  e^{i (k_1-k_2)x}\right|^2 ,
\end{equation}
where $w(x,k)$ represents the source function which describes
the mean number of particles of four-momentum $k$ emitted from
a source element centered at the space-time point $x$.
In order to examine the correlation function in longitudinal
and transverse direction it is convenient to introduce
the quantities
$k=(k_\perp\cosh y,k_\perp\cos \psi,k_\perp\sin \psi,k_\perp\sinh y)$,
$K = \frac{1}{2} (k_1+k_2) = (K_0,K_l,\vec{K}_\perp)$ and $
q =  k_1-k_2 = (q_0,q_l,\vec{q}_\perp))$
and the projection $q_{out}$, which is the
component of $\vec{q}_\perp $ parallel
to $\vec{K}_\perp$.
The analysis of the BECs in the direction of $q_l,q_{out}$
reflects properties of the longitudinal and  transverse
space-time expansion
of the system.

Fig. 2a shows the
Bose-Einstein correlation function in longitudinal direction
for thermal photons produced
in $Au+Au$ collisions at $\sqrt{s}=200$ AGeV.
It is calculated for the momentum configuration $k_{1\perp}=k_{2\perp}$,
$\psi_1=\psi_2=0$ and $y_1=0$ as a function of the rapidity difference of
the two photons $\Delta y=y_2-y_1$.
 In  Fig. 2b the correlation function is plotted as a function of
$q_{out}=q_\perp \frac{\vec{k}_\perp\cdot\vec{q}_\perp}{k_\perp q_\perp}$.
We consider the case $y_1=y_2=0, \psi_1=\psi_2=0$.
Figs. 2a and 2b reveal that the shapes and the widths
of the correlation functions change drastically as
$k_\perp$ increases from $1$ to $2$ GeV.
At 1 GeV the BEC functions have an exponential shape and a small width
whereas
the shape turns into a gaussian and the width increases strongly as
 $k_\perp$ reaches values of $2$ GeV and higher.

This characteristic behaviour is due to the fact that
low transverse momentum ($k_\perp\sim1$ GeV) photons are produced both
in the pre-equilibrium and the hadronic phase (see Fig. 1).
Photons  with $k_\perp>1.5$ GeV, on the other hand, are
almost exclusively
emitted from
the pre-equilibrium stage.
This is reflected in the widths of the photon correlation
 functions  (Fig. 2a and 2b).
which are inversely proportional to the size of
the emission region.
The  broadest correlation function
($k_\perp=5$ GeV) is related to the hot pre-equilibrium stage which
has minimum spatial extension whereas the correlation function for
pairs of lower momenta is dominated by the late hadronic stage
which has
a large spatial extension.

Thus, the rapid change of the shape of the correlation functions in the region
$k_\perp=1-2$ GeV is a direct consequence of the fact that at $k_\perp
\sim 1$ GeV the contribution of the hadronic phase to the photon yield
becomes comparable to the contribution of the pre-equilibrium phase (which
dominates above $k_\perp \sim 1.5$ GeV), i.e., that for this momentum
range there are significant contributions from two separate sources.

One may ask how it is possible for the
hadronic phase to emit $1$ GeV photons at a rate comparable to the
pre-equilibrium contribution while the contributions from the equilibrium
QGP remain smaller by about an order magnitude. The reason is that at a
given temperature the emission rates from a hadronic resonance
gas are considerably larger than the rates from a system of
quasi-free quarks and gluons.
The larger rates from the hadron gas are mainly due to the contribution
the $A_1$ meson \cite{xiong} which were not taken into account
in previous studies
\cite{ich,longevity,history,fidelity}
of BEC's of thermal photons.
Thus the change in the shape of the correlation function is indeed
due to a change in the production mechanism of photons between quark-matter
and the hadronic resonance gas. This implies that the transition from a
hadron gas to a deconfined
quark-gluon stage can be seen by photon-interferometry.

In Fig. 3 we show
how the correlation function would behave in the absence of this
change of production mechanism associated with the phase transition.
To be specific,
we have taken the equation of state for a resonance gas and connected
it to the PCM solution at t=2.4 fm/c. For the entire ``thermal'' history
(including the pre-equilibrium with its effective temperatures)
we used the photon rates obtained for
the hadron gas in Ref. \cite{xiong}. As far as the early pre-equilibrium
stages are concerned, this is of course not a realistic scenario. The
only reason we discuss it here is to demonstrate explicitly that the
change in shape of the correlation function observed in Figs. 2a and 2b
is in fact a consequence of the different production mechanisms in the
hadronic and the quark-gluon phase. Indeed, as can be seen in Fig. 3
for the ``purely hadronic'' scenario the correlation function does not
exhibit a pronounced $k_\perp$ dependence anymore. Moreover, the
BEC function does not have the two-component structure characteristic for the
contribution from two separate phases. Rather, the curves all have
an approximate Gaussian shape.

{} From our hybrid model, which connects the PCM with
relativistic 3d-hydrodynamics and includes photon production in the early
pre-equilibrium stage, we conclude that photon interferometry
in the transverse momentum region between 1 and 3 GeV can serve as a probe of
the change in the production mechanism of thermal photons associated with
the transition from a system of quasi-free quarks and gluons to a
hadronic resonance gas.
Whereas
the inclusive one-photon spectrum is dominated by
the extremly "hot" pre-equilibrium phase it turns out that the
two-photon correlation function exhibits
a strong transverse momentum dependence which is a signature
of the deconfinement phase transition.\\[3ex]

\noindent
This work was supported in part by the Federal Minister of Research and
Technology under contract 06MR731 and the Gesellschaft f\"ur
Schwerionenforschung (GSI).
R.W. acknowledges the support of the Deutsche
Forschungsgemeinschaft and the hospitality of A. Capella, LPTHE,
Univ. de Paris-Sud during the last stages of this study.

\newpage

{\Large \bf Figure Captions}\\
\begin{description}
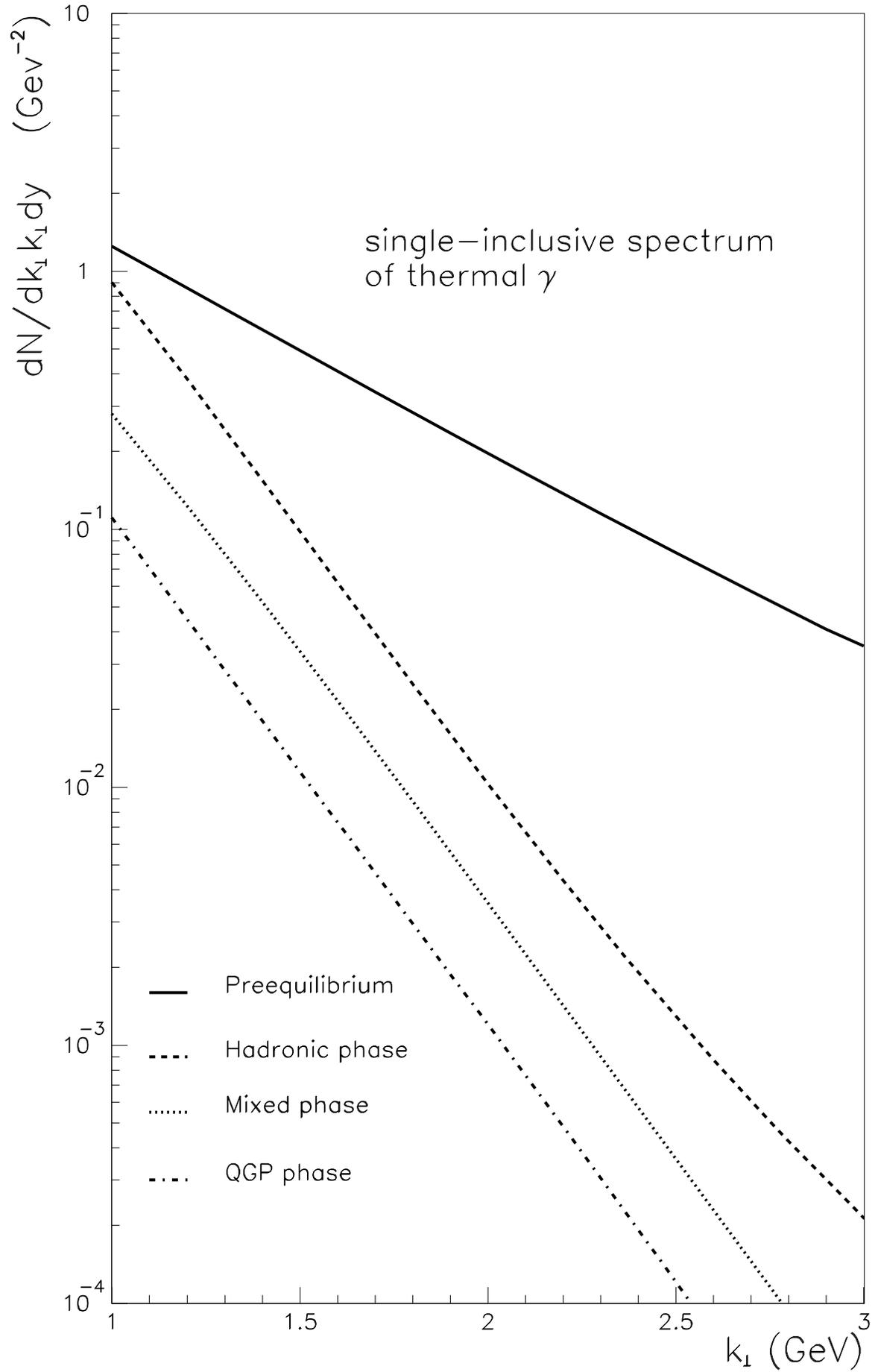
\item[Fig. 1]  Single-inclusive spectrum of pre-equilibrium
and thermal photons for $Au+Au$ collisions at $E_{lab}=200$ AGeV
as a function of
the transverse momentum
at rapidity y=0.
The contributions of the different phases are displayed seperately.
\item[Fig. 2a]  Bose-Einstein correlations of photons as
a function of the rapidity difference of the photons
calculated for different values of the transverse momentum $k_{\perp}$.
\item[Fig. 2b]  Bose-Einstein correlations of photons as
a function of the variable $q_{out}$
calculated for different transverse momenta $k_{\perp}$
of one photon.
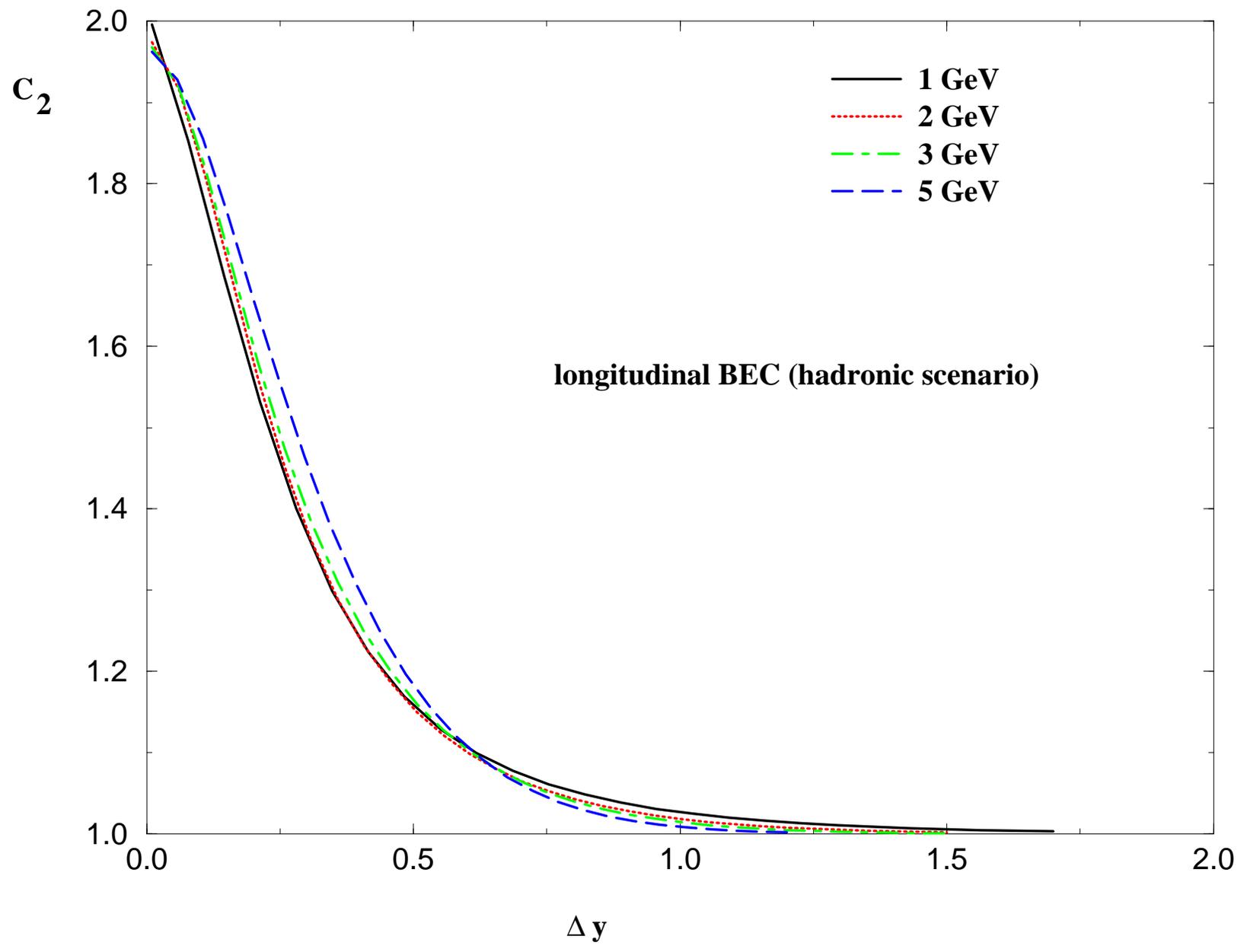
\item[Fig. 3]  Bose-Einstein correlations of photons as
a function of the rapidity-difference of the photons
calculated for different transverse momenta $k_{\perp}$
of one photon. For the calculation we took a purely hadronic
scenario.\\

\end{description}

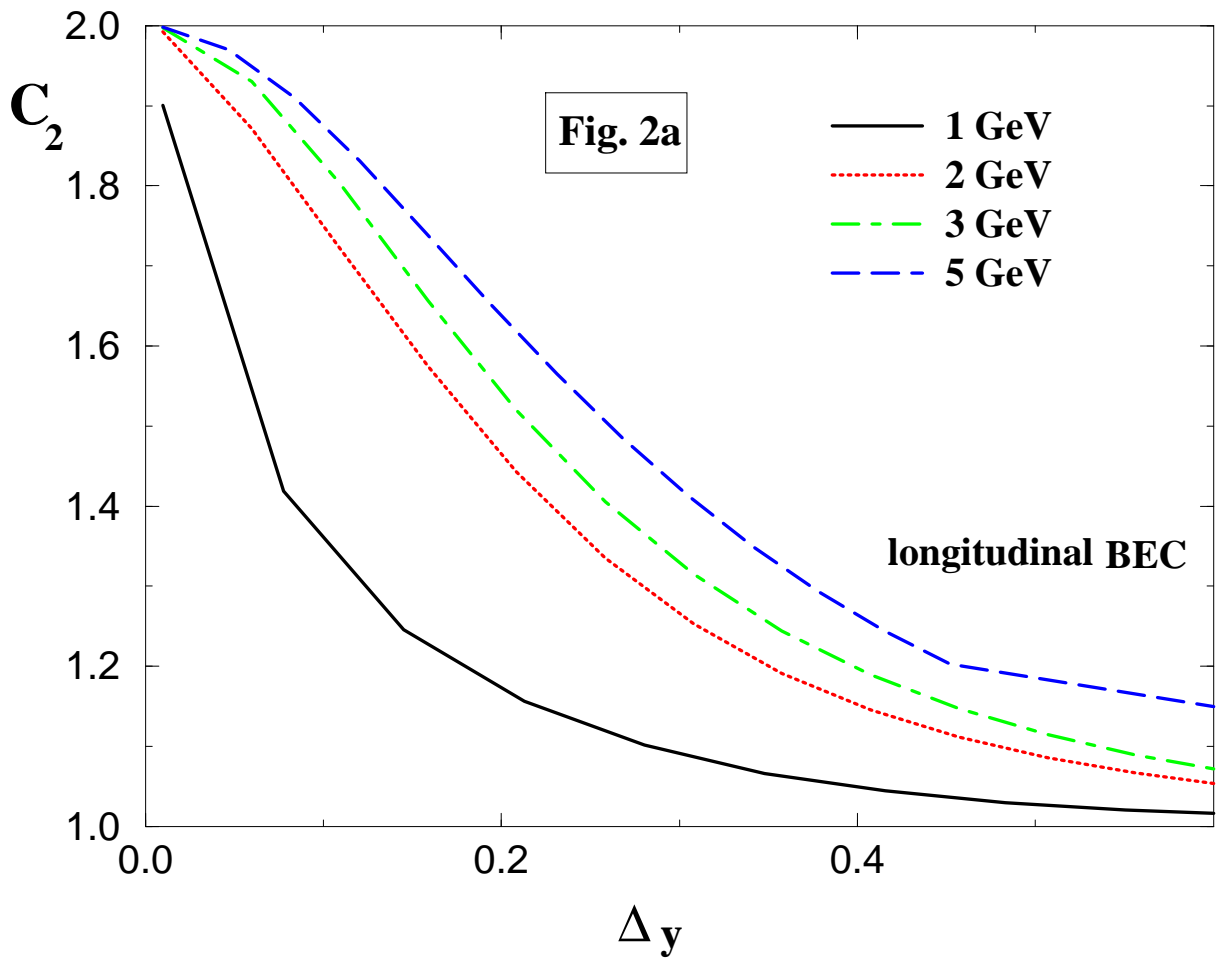
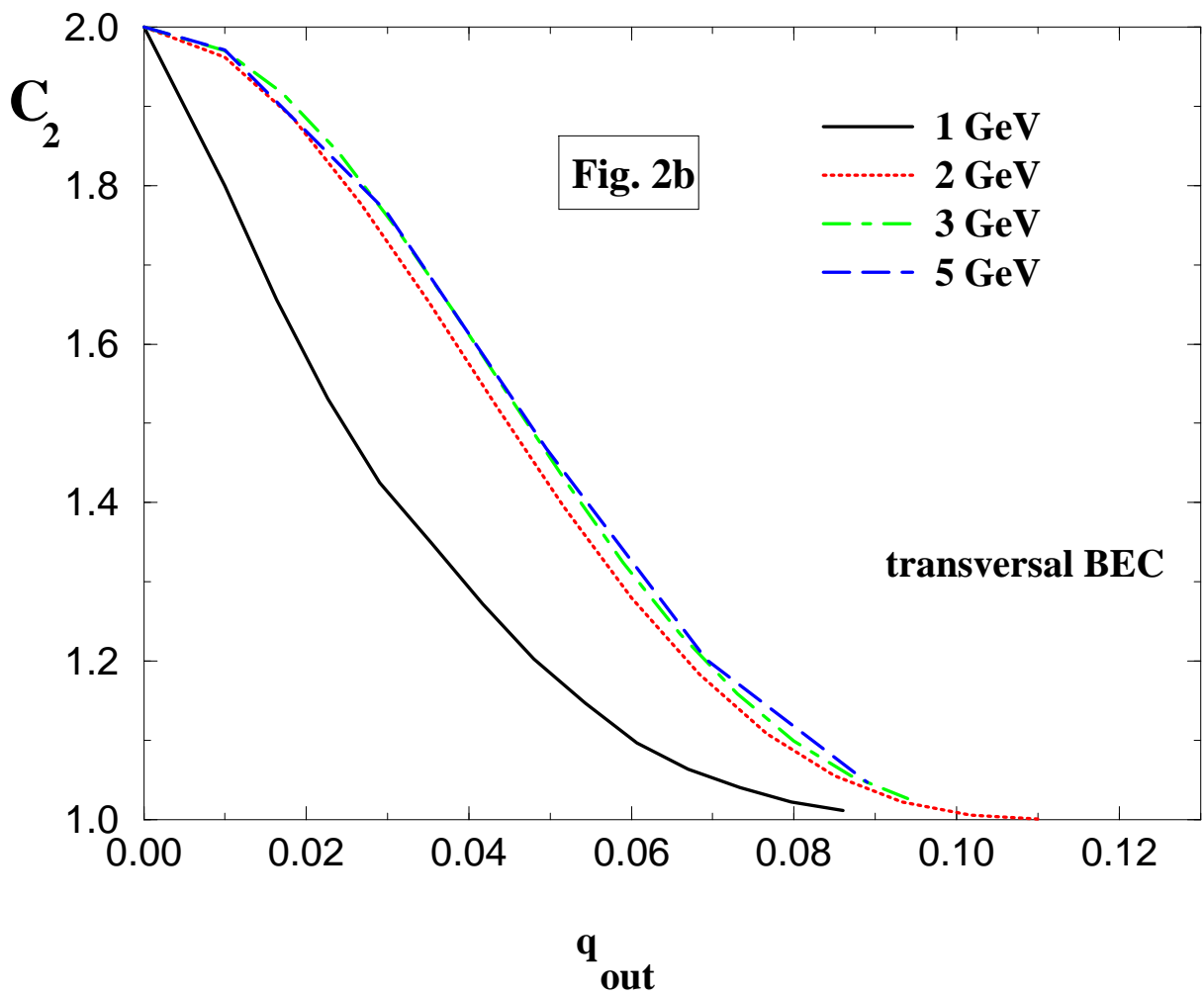
\end{document}